\documentclass[12pt,preprint]{aastex}

\received{}
\revised{}
\accepted{}

\cpright{PD}{2006}

\renewcommand{\d}{\mathrm{d}}
\newcommand{\pubjournal}[4]{#4, #1, #2, #3}

\newcommand{\eVdist}{\kern-0.06667em}
\newcommand{\Pev}{{\mathrm{Pe}\eVdist\mathrm{V\/}}}
\newcommand{\pev}{{\,\mathrm{Pe}\eVdist\mathrm{V\/}}}
\newcommand{\Tev}{{\mathrm{Te}\eVdist\mathrm{V\/}}}
\newcommand{\tev}{{\,\mathrm{Te}\eVdist\mathrm{V\/}}}
\newcommand{\Gev}{{\mathrm{Ge}\eVdist\mathrm{V\/}}}
\newcommand{\gev}{{\,\mathrm{Ge}\eVdist\mathrm{V\/}}}
\newcommand{\cm}{{\,\mathrm{cm}}}
\newcommand{\km}{{\,\mathrm{km}}}

\newcommand{\scnd}{{\,\mathrm{s}}}

\begin{document}

  \title{Potential Neutrino Signals from Galactic $\gamma$-Ray Sources}
  
\author{Alexander Kappes}
\affil{Friedrich-Alexander-University Erlangen-Nuremberg, Erlangen, Germany}
\email{kappes@physik.uni-erlangen.de}
\author{Jim Hinton}
\affil{School of Physics and Astronomy, The University of Leeds, Leeds, UK}
\affil{Max-Planck-Institut f\"ur Kernphysik, Heidelberg, Germany}
\affil{Landessternwarte, Universit\"at Heidelberg, K\"onigstuhl, Heidelberg, Germany}
\author{Christian Stegmann}
\affil{Friedrich-Alexander-University Erlangen-Nuremberg, Erlangen, Germany}
\author{Felix A. Aharonian}
\affil{Dublin Institute for Advanced Studies, Dublin, Ireland}
\affil{Max-Planck-Institut f\"ur Kernphysik, Heidelberg, Germany}

\begin{abstract}
  The recent progress made in Galactic $\gamma$-ray astronomy using
  the High Energy Stereoscopic System (H.E.S.S.) instrument provides
  for the first time a population of Galactic $\Tev$ $\gamma$-ray, and
  hence potential neutrino sources, for which the neutrino flux can be
  estimated. Using the energy spectra and source morphologies measured
  by H.E.S.S., together with new parameterisations of pion production
  and decay in hadronic interactions, we estimate the signal and
  background rates expected for these sources in a first-generation
  water Cherenkov detector (ANTARES) and a next generation neutrino
  telescope in the Mediterranean Sea, KM3NeT, with an instrumented
  volume of $1\km^3$. We find that the brightest $\gamma$-ray sources
  produce neutrino rates above $1\tev$, comparable to the background
  from atmospheric neutrinos. The expected event rates of the
  brightest sources in the ANTARES detector make a detection unlikely.
  However, for a $1\km^{3}$ KM3NeT detector, event rates of a few
  neutrinos per year from these sources are expected, and the
  detection of individual sources seems possible. Although generally
  these estimates should be taken as flux upper limits, we discuss the
  conditions and type of $\gamma$-ray sources for which the neutrino
  flux predictions can be considered robust.
\end{abstract}


%
%

\section{Introduction}

Neutrinos and $\gamma$-rays in the $\Gev$--$\Pev$ range provide the
only available probes of cosmic ray (CR) acceleration sites and CR
propagation in our Galaxy. Both particles are produced in hadronic
interactions of cosmic protons and nuclei with the ambient gas and are
not deflected in the interstellar magnetic field. Despite many years
of research, no source of high-energy cosmic neutrinos has yet been
identified~\citep{AMANDA, BAIKAL}. This experimental situation is
unfortunate, as the detection of even a single cosmic $\Tev$ neutrino
could potentially provide compelling evidence for the long
sought-after class of hadronic CR accelerators.  On the other hand,
$\Tev$ $\gamma$-ray astronomy has recently made considerable progress
with the detection of several shell-type supernova remnants (SNRs),
the prime candidates for the acceleration of Galactic CRs (see, e.g.,
\citet{Hillas_snr} for a recent review). As $\Tev$ neutrinos should be
produced in roughly equal numbers to $\Tev$ photons in any hadron
accelerator, these objects represent prime targets for neutrino
telescopes.  Moreover, the recent High Energy Stereoscopic System
(H.E.S.S.)~\citep{HESS} survey of the Galactic plane
\citep{HESS_scan,HESS_scan2} has revealed several additional classes
of potential high energy Galactic neutrino sources: X-ray binaries,
pulsar wind nebulae (PWNe), and unidentified but probable
hadron-accelerating sources such as HESS\,J1303$-$631
\citep{HESS_1303}, as well as diffuse emission from the interactions
of CRs in the Galactic center (GC) region
\citep{HESS_gc}.
 
The case for CR acceleration in SNRs was strengthened by the detection
of resolved $\Tev$ $\gamma$-ray emission from the shells of two SNRs
\citep{HESS_1713,HESS_velajnr}. However, while these measurements
provide direct evidence of particle acceleration in these objects, in
both cases an ambiguity remains between electrons and hadrons as the
radiating particles. Both inverse Compton scattering of relativistic
electrons and decay of neutral pions produced in hadronic
interactions, can satisfactorily explain the available data. In fact,
the only instance in which a very strong case for very high-energy
(VHE) $\pi^{0}$-decay emission can be made is for the giant molecular
clouds of the GC region, due to the strong spatial correlation between
$\gamma$-ray emission and available target material (interstellar
gas). However, even in this case the nature of the CR source remains
uncertain, as at least two good candidates exist: the SNR Sgr\,A\,East
\citep{Crocker_saga} and the black hole
Sgr\,A$^{\star}$~\citep{Aharonian_saga}. The case for hadron
acceleration in unidentified $\gamma$-ray sources can be made based on
a lack of associated X-ray emission, which would be expected for cosmic
\emph{electron} accelerators. 

Some of the brightest sources in the $\Tev$ $\gamma$-ray sky are PWNe,
for example, the nebulae associated with the powerful Crab and Vela
pulsars~\citep{HEGRA_Crab, HESS_velaX}. The $\gamma$-ray emission from
these objects is normally interpreted as inverse Compton upscattering
of cosmic microwave background radiation (or synchrotron) photons by
energetic electrons. However, the existence of a significant fraction
of nuclei in pulsar winds has been suggested~\citep{Hoshino_PWN}. In
some scenarios, the $\Tev$ emission may be dominated by the decay of
pions produced in the interactions of these nuclei, and significant
production of neutrinos may occur (see, e.g.,
\citet{horns_velax,Bednarek_PWN,amato_pwn}).

Finally, $\Tev$ $\gamma$-ray emission has recently been reported from
the X-ray binaries LS~5039
\citep{HESS_ls5039} and LS\,I\,+61~303 \citep{MAGIC_lsi61303}. The
rapid cooling of $\Tev$ electrons in the dense radiation fields
associated with such objects is suggestive of a hadronic origin for
the $\gamma$-radiation. While the $\gamma$-ray emission from these
sources is relatively weak, they are promising candidates for neutrino
emission because of the strong $\gamma$-ray absorption that likely
occurs within the sources~\citep{Aharonian_ls5039}.

Given these recent advances in our knowledge of the $\Tev$
$\gamma$-ray emission of Galactic sources, and the planned or current
construction of major $\Tev$ neutrino detectors, we consider this an
appropriate time to consider in detail the detectability of neutrino
counterparts to these sources in the medium term. Since H.E.S.S.\
measures $\gamma$-ray spectra up to $10\tev$ and in some cases even
above $10\tev$, the prediction for $\Tev$ neutrino signals can be done
quite robustly, especially for extended sources for which the internal
$\gamma$-ray absorption is negligible.

In contrast to the search for extra-Galactic neutrino sources, the
geographic location of the detector is of vital importance to a search
for Galactic hadron accelerators. In the northern hemisphere,
Cassiopeia\,A is the only known $\gamma$-ray SNR and has a rather weak
$\Tev$ flux ($\sim2\times10^{-12} \, \mathrm{erg} \, \mathrm{cm}^{-2}
\, \mathrm{s}^{-1}$, 1\,--\,$10\tev$) \citep{HEGRA_casa}. 
The unidentified northern hemisphere source $\Tev$\,J2032+4130 is also
rather weak \citep{HEGRA_TeV}, and for the extended $\Tev$ source
reported in \citet{milagro_hotspot}, the spectrum is unknown so far. On the
other hand, the strongest northern hemisphere $\gamma$-ray source, the
Crab Nebula, with a measured energy spectrum exceeding $50\tev$, is
very likely an electron accelerator~\citep{HEGRA_Crab}. This leaves
the recently detected $\gamma$-ray emitter, the X-ray binary
LS\,I\,+61~303, as the only possible strong Galactic neutrino source
so far, providing that the $\gamma$-ray emission is caused by pion
decay and that strong $\gamma$-ray absorption occurs.

\begin{figure*}[t]
\begin{center}
\plotone{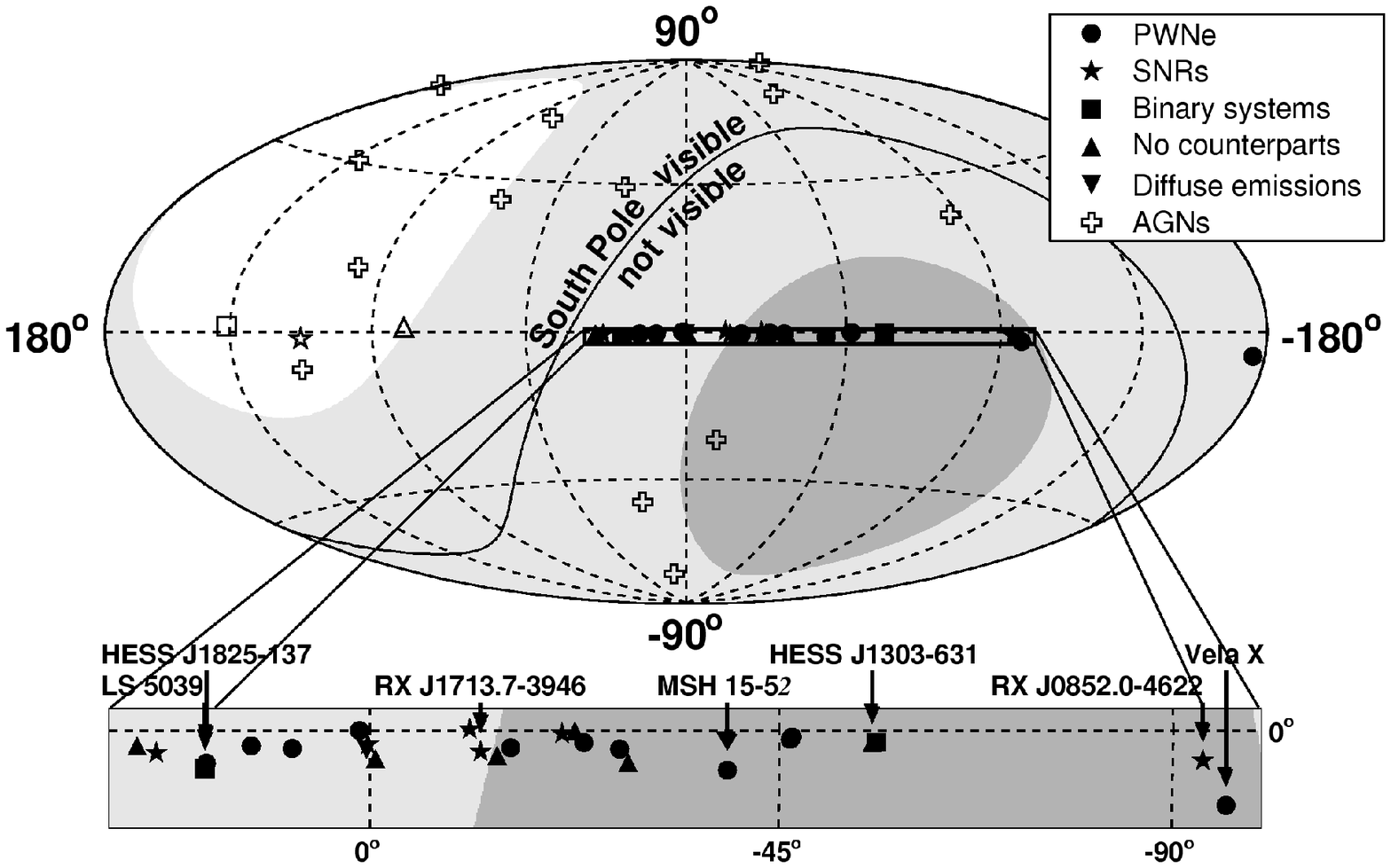}
\caption{ 
  Sky map of $\Tev$ $\gamma$-ray sources in Galactic coordinates
  together with their visibility to neutrino telescopes. Source types
  are represented by different symbols: filled symbols mark sources
  discussed in this paper, the shaded regions represent 25\%\,--\,75\%
  (\emph{light gray}) and $>75$\% (\emph{dark gray}) visibility to a
  detector in the central Mediterranean Sea for energies below $\sim
  100 \tev$, and the solid curve shows the extent of the visibility of
  a South Pole detector (IceCube). The inset displays a zoomed view of
  the area between $-100^\circ$ and $+28^\circ$ in Galactic longitude
  and about $-3.5^\circ$ and $+0.8^\circ$ in Galactic latitude. }
\epsscale{1.0}
\label{fig:skymap}
\end{center}
\end{figure*}

The situation in the southern hemisphere is strikingly different: at
least five SNRs have now been detected, and two, RX\,J1713.7$-$3946
and RX\,J0852.0$-$4622, are extremely bright ($\sim 10^{-10} \,
\mathrm{erg} \, \mathrm{cm}^{-2} \, \mathrm{s}^{-1}$, $0.5$\,--\,$10 \tev$)
\citep{HESS_1713,HESS_velajnr}. In addition, the southern hemisphere
contains the unique collection of objects at the GC and many as yet
unidentified sources. The probability of detection of a Galactic
neutrino source by a (downward-looking) Northern Hemisphere detector
therefore appears much greater than for a Southern Hemisphere
instrument. In this paper we discuss in detail the observability of
neutrino emissions from CR accelerators in the ANTARES detector
\citep{ANTARES_misc1}, currently under construction in the
Mediterranean Sea, and an additional planned km$^3$ scale detector in
the Mediterranean Sea, KM3NeT \citep{KM3NET} (in this paper we assume
for KM3NeT an instrumented volume of $1
\km^3$). Figure~\ref{fig:skymap} shows the sky map of known $\Tev$
$\gamma$-ray sources and their visibility to the major Northern and
Southern Hemisphere neutrino telescopes.
Since it is not possible with current neutrino telescopes to separate
neutrinos and anti-neutrinos the word neutrino is used to encompass
both throughout the article.

%
%
\section{Galactic $\gamma$-ray emission}

\subsection{Resolved sources}

The H.E.S.S.\ survey of the inner Galaxy~\citep{HESS_scan2}, taken
together with targeted observations of southern hemisphere Galactic
sources, provides a comprehensive list of VHE $\gamma$-ray sources
relevant to the current work. Earlier claims of $\gamma$-ray emission
from southern Galactic sources have either been confirmed (e.g.,
\citet{HESS_1713}) or refuted (e.g., \citet{HESS_sn1006}).
Table~\ref{tab:rates} lists all 27 southern hemisphere Galactic
sources reported so far by the H.E.S.S.\ collaboration. Based on
information in the H.E.S.S.\ publications and other relevant
multi-wavelength information, we assign each source into one of four
categories: (A) Unambiguously associated with a SNR shell, (B)
associated with a binary system, (C) lacking any good counterpart at
other wavelengths and (D) plausibly associated with a PWN.  While the
assignment of some objects is rather arbitrary (particularly between
categories C and D), and some assignments will undoubtedly change as
better multi-wavelength data appear, we nevertheless consider that the
objects of category A, and perhaps also category C, are most likely
candidates for neutrino emission. The most promising objects are, of
course, young shell-type SNRs. The morphological and spectrometric
characteristics of $\Tev$ $\gamma$-ray emission from two objects of
this category, RX\,J1713.7$-$3946 and RX\,J0852.0$-$4622, have been
studied by H.E.S.S.\ with great detail.  At $1\tev$, both sources show
fluxes as large as the Crab Nebula flux but with significantly harder
energy spectra extending beyond $10\tev$. If the hadronic
$\gamma$-rays from interactions of CR with the ambient matter dominate
over the inverse Compton component, then the neutrino fluxes from the
shells of these objects can be calculated with a good accuracy in the
most relevant energy band between $0.1$ and $100\tev$.

For some objects with angular sizes less than $\sim0.2^{\circ}$, an
affiliation with a particular category is ambiguous as a
discrimination between shell-like emission and that of any central
nebula is difficult. HESS\,J1813$-$178 is one such object, with a
shell-like radio emission \citep{VLA_1813}, a central X-ray nebula
\citep{ASCA_1813}, and an unresolved $\Tev$ $\gamma$-ray emission
\citep{HESS_scan2}.

Category B and D objects are generally treated as leptonic (inverse
Compton) sources and therefore are less likely neutrino emitters.
However, interpretations of $\Tev$ $\gamma$-ray emission from two
representatives of these classes, namely, the microquasar LS~5039 and
the plerion Vela\,X, in terms of hadronic interactions are quite
possible (see, e.g., \citet{Aharonian_ls5039,horns_velax}). In this
paper, neutrino fluxes are calculated for all objects in classes A, B,
C and D.

\subsection{Undetected hard-spectrum sources}

Experiments such as H.E.S.S.\ have substantially reduced energy flux
sensitivities beyond $\sim 10 \tev$ relative to their $1 \tev$
performance. For this reason bright sources with extremely hard
spectra between $\sim 100 \tev$ and $1 \pev$ could in principle be
``missed'' by current Cherenkov telescopes. Due to the rapid rise of
the effective collection area of neutrino telescopes with energy, such
sources could be promising candidates for these detectors. It is
instructive to consider how hard the energy spectrum of a source must
be to be undetected in $\Tev$ $\gamma$-rays and potentially detectable
in $\sim 100 \tev$ neutrinos. The H.E.S.S.\ survey of the Galactic
plane reaches a typical sensitivity of $1
\times 10^{-13} \cm^{-2} \scnd^{-1}$ above $10 \tev$. To produce a
detectable flux in a $\km^{3}$ volume neutrino detector and be missed
by this survey, we estimate that a photon index of at least about 1 is
required. In addition to the Cherenkov telescope limits, air-shower
arrays have produced limits on ultra-high-energy $\gamma$-ray fluxes
for most of the sky. The CASA-MIA detector was used to produce limits
on northern hemisphere point sources of $\sim4\times10^{-14} \cm^{-2}
\scnd^{-1}$ above $140\tev$~\citep{CASA}. The SPASE detector at the
South Pole produced limits over the region $\delta < -45^{\circ}$ of
$\sim2\times10^{-13}
\cm^{-2} \scnd^{-1}$ above $50 \tev$~\citep{SPASE}. Updated results
from the larger area and longer integration time of the SPASE-2
instrument~\citep{SPASE2} may be useful in constraining the ($>50
\tev$) extrapolation of the spectra of strong southern hemisphere
sources such as RX\,J0852.0$-$4622 in the future.

\subsection{Diffuse emission and unresolved sources}

For instruments with modest angular resolution and a very wide field
of view, it may be easier to detect large-scale diffuse emission than
to resolve individual sources. This fact is demonstrated by the recent
MILAGRO detection of a diffuse $\Tev$ $\gamma$-ray signal from the
Galactic plane between $l=40^{\circ}$ and $100^{\circ}$
\citep{MILAGRO}. Any such signal must be considered as the sum of the
emission induced by CR (hadron and electron) interactions with a
contribution from unresolved sources. The calculation of diffuse
neutrino and $\gamma$-ray emission arising from CR interactions in our
Galaxy is rather complex (see, e.g., \citet{GALPROP}). However, at
least for the emission from CR hadrons, a reasonable approximation
simplifies the situation enormously: for constant emissivity, the
$\gamma$-ray and neutrino signals are simply proportional to the
product of the total column depth of material (molecular + atomic)
with the mean CR density in the volume integrated.  Following
~\citet{AharonianClouds} the emissivity for CRs is such that
\begin{eqnarray}
F_{\gamma}(> E_{\gamma}) &=& 4 \times 10^{-14} S
\left(\frac{E_{\gamma}}{1\tev}\right)^{-1.7} 
\mathrm{cm}^{-2}\,\mathrm{s}^{-1} \\
\mbox{with}\quad S &=& \left(\frac{\Omega}{1\,\mathrm{deg}^{2}}\right)
\left(\frac{N_{H}}{10^{22}\,\mathrm{cm}^{-2}}\right) \ ,
\end{eqnarray}
assuming that the locally measured CR spectrum is valid throughout the
Galaxy. While this assumption may be valid in regions far from active
CR accelerators, enhanced emission is expected in the neighbourhood of
such sources. The relatively high $\gamma$-ray flux detected from the
giant molecular clouds of the GC region \citep{HESS_gc}
implies an enhanced CR density in the central 200\,pc of our
Galaxy. However, this GC emission is still rather weak ($10^{-11}
\,\mathrm{erg}\scnd^{-1}$ in the 0.2\,--\,$20\tev$ range), making a
detection of the associated neutrino emission rather difficult. The
integrated diffuse emission from the entire Inner Galaxy (diffuse
plane emission) may present a more promising target. In the window
$|b|<1^{\circ}$, $|l|<30^{\circ}$, the approximate mean molecular and
atomic column densities are $4 \times 10^{22}$ and $1.5
\times 10^{22} \cm^{-2}$, respectively \citep{DameCO}. Assuming an
enhancement factor of 1.4 of the CR density in the inner Galaxy
relative to the local density~\citep{HunterCR}, the predicted diffuse
flux is $\sim 3.5 \times 10^{-11} \cm^{-2} \scnd^{-1}$ ($> 1\tev$),
with a spectral index close to that of the parent spectrum, i.e.,
$\sim 2.7$. As this calculation assumes no contribution from electrons
above $1\tev$ (as suggested by more detailed models from
~\citep{GALPROP}), this flux is valid for both $\gamma$-rays and
neutrinos.

An estimate of the contribution from unresolved sources requires a
population model. While the number of detected sources is relatively
small, a meaningful comparison with simple population models is
possible. We consider here a model (discussed also
in~\citet{HESS_scan2}) in which mono-luminous sources are distributed
within the Galaxy roughly as the distribution of molecular material.
This simple approach is sufficient to explain the latitudinal and flux
distributions of the detected sources.  This simple model suggests
that the unresolved component represents $\sim 30\%$\,--\,$50\%$ of
the total flux. Equally uncertain is the fraction of the detected
sources in which neutral pion decay is the dominant $\gamma$-ray
production mechanism. The fraction of the total \emph{detected}
$\gamma$-ray flux represented by source classes A, B and C in the
region $|b|<1^{\circ}$, $|l|<1^{\circ}$ is $\sim50$\% (a sub-area of
this region with $|b|<0.3^\circ$ and $|l|<0.8^\circ$ is later referred
to as the ``Galactic center ridge''). Diffuse $\gamma$-ray and
neutrino fluxes above $1 \tev$ from this region are therefore $\sim
3\times10^{-11}$ and $\sim 1.5\times10^{-11} \cm^{-2} \scnd^{-1}$,
respectively. The spectral index of the unresolved class is most
likely close to that of the detected sources, i.e., $\sim 2.3$.  The
harder spectral index of the source component makes any neutrino
detection of diffuse emission likely to be dominated by unresolved
sources.

Applying the same method to the region
$40^{\circ}\,<\,l\,<\,100^{\circ}$, $|b|\,<\,5^{\circ}$ results in a
flux of $4\times10^{-10} \cm^{-2} \scnd^{-1}$, compatible with the
MILAGRO measurement in this region: $F(>1\tev)$ = $(5.1 \pm 1.0 \pm
1.7)\times 10^{-10}\cm^{-2} \scnd^{-1}
\,\mathrm{sr}^{-1}$~\citep{MILAGRO}. The expected contribution from
unresolved sources is $\sim 60$\% in this case. As the region is not
very visible for Mediterranean detectors it is not discussed further
in this paper.

%
%
\section{Neutrino fluxes}

In $\gamma$-ray sources where the $\Tev$ emission is dominated by the
decay of $\pi^{0}$ particles produced in $p$-$p$ interactions, the
measured $\gamma$-ray spectra can be used to derive the expected
$\Tev$ neutrino spectra. Several calculations of expected neutrino
spectra have been made for the special case of a power-law spectrum of
primary hadrons
\citep{berezinsky,gaisser,crocker,halzen,costantini,lipari}. However,
given that for many sources curved spectra are measured (or expected)
in the energy range of neutrino detectors ($E \approx 1\tev$ to $E >
1\pev)$, a revised treatment of such spectra is required.

A recent parameterisation of the pion and secondary particle
production in hadronic interactions (based on results from the SIBYLL
event generator; \citet{SIBYLL}) can be used to derive the $\gamma$,
$\nu_{e}$, and $\nu_{\mu}$ spectra for arbitrary incident proton
spectra~\citep{kelner}. We have used these parameterisations to
calculate the relationship between $\gamma$-ray and neutrino spectra
in the case in which the primary proton spectrum is given by a
power-law with index $\alpha$ and an exponential cut-off energy
$\epsilon_p$:
\begin{eqnarray}
\frac{\d N_{p}}{\d E_{p}} =
k_{p}\,\left(\frac{E_{p}}{1\tev}\right)^{-\alpha}\,\exp\left(-\frac{E_{p}}
{\epsilon_{p}}\right)
\label{eq1}
\end{eqnarray}
for protons with energy $E_{p}$.

Based on the parameterisations we calculated the resulting summed
electron and muon neutrino spectrum $\d N_\nu/\d E_\nu$ at the source
($\nu_e$\,:\,$\nu_\mu = 1:2$). Assuming full mixing the muon neutrino
spectrum at Earth is then given by one-third of $\d N_\nu/\d
E_\nu$. The resulting muon neutrino spectrum is only a little
dependent on the relative fraction of the electron and muon neutrinos
at the source.

\begin{figure}[t]
\begin{center}
\plotone{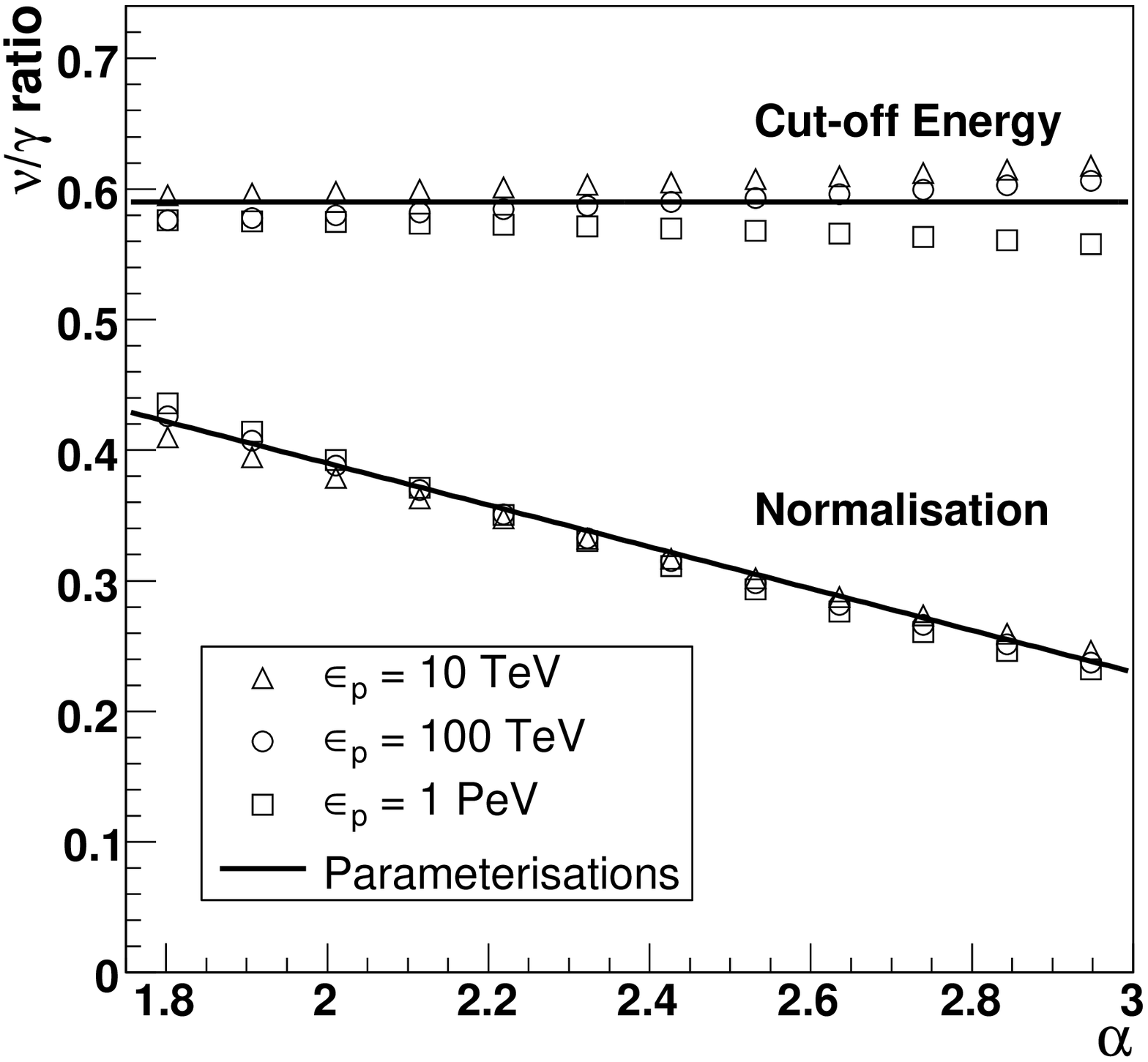}
\caption{ Ratio of the best-fit neutrino and $\gamma$-ray normalisations
  ($k_{\nu}/k_{\gamma}$) and cut-off energies
  ($\epsilon_{\nu}/\epsilon_{\gamma}$) for a range of proton indices
  $\alpha$ and cutoff energies $\epsilon_p$.  The indices of the
  secondary spectra are kept fixed at $\alpha - 0.1$. The curves show
  the parameterisations given in the text.}
\label{fig:parameterisation}
\end{center}
\end{figure}

Considering spectra with $1.8<\alpha<3.0$ and $10\tev <\epsilon_{p} <1
\pev$ and assuming full neutrino mixing, we find that the spectra of
$\gamma$-rays and muon neutrinos at the Earth can be described by
\begin{eqnarray}
\frac{\d N_{\gamma/\nu}}{\d E_{\gamma/\nu}} \approx k_{\gamma/\nu}
\left(\frac{E_{\gamma/\nu}}{\mathrm{1\tev}}\right)^{-\Gamma_{\gamma/\nu}}\,
\exp\left(-\sqrt{\frac{E_{\gamma/\nu}}{\epsilon_{\gamma/\nu}}}\right)
\label{eq2}
\end{eqnarray}
and the parameters $k$, $\Gamma$, and $\epsilon$ are given by
\begin{eqnarray}
k_{\nu}&\approx& (0.71 - 0.16\alpha)\,k_{\gamma}\\
\Gamma_{\nu} &\approx& \Gamma_{\gamma}\,\approx\,\alpha - 0.1\\
\epsilon_{\nu}&\approx&
0.59\,\epsilon_{\gamma}\,\approx\,\epsilon_{p}/40 \ .
\end{eqnarray}
Figure~\ref{fig:parameterisation} illustrates the accuracy of the
parameterisations of the cut-off energy and the
normalisation. Equation\,\ref{eq2} provides a satisfactory fit to the
$\gamma$-ray spectra of all sources detected using H.E.S.S., where for
sources with no published claim of a curvature a pure power law is
fitted ($\epsilon_\gamma = \infty$). By refitting the spectral points
(for references see Tab.\,\ref{tab:rates}) to this equation we produce
predicted muon neutrino spectra under the assumptions given above.

Figure~\ref{spectra} shows in the top plots as examples the neutrino
spectra of two of the strongest Galactic $\gamma$-ray sources with
observed cutoff, the SNR RX\,J1713.7$-$3946 and the PWN Vela\,X. The
open and shaded areas of the $\gamma$-ray and neutrino spectra,
respectively, include the $1 \,\sigma$ systematic uncertainties. This
we assume to be 20\% on the normalisation, 10\% on the cutoff energy,
and 0.1 on the index based on the systematic uncertainties of the
published H.E.S.S.\ $\gamma$-ray spectra.  Table~\ref{tab:rates}
contains the parameters of the neutrino fluxes for all sources, where
it should be noted that $\Gamma_\nu$ and $\epsilon_\nu$ are strongly
correlated.

Some essential assumptions are made in the calculation of the
predicted signal, and it is important to assess the validity of these
assumptions on a source-by-source basis. The most important of these
assumptions are

\begin{enumerate}
\item no significant contribution of non hadronic processes to the
  measured $\gamma$-ray signal;
\item no significant $\gamma$-ray absorption within the
  source, i.e.\ radiation and matter densities are sufficiently low
  for most $\gamma$s to escape;
\item no significant $p\gamma$ interaction (radiation
density low);
\item charged pions decay before interacting (matter density is low);
\item muons decay without significant energy loss (magnetic field is low);
\item nucleus-nucleus interactions produce pion spectra that are similar
  enough to the $p$-$p$ case that they can be treated in the same way;
\item the size of the emitting region within each source is large
   enough that oscillations will produce a fully mixed neutrino 
   signal at the Earth ($\nu_{e}:\nu_{\mu}:\nu_{\tau} = 1:1:1$).
\end{enumerate}

For all of the extended $\gamma$-ray sources detected by H.E.S.S., it
seems likely that these conditions (with the likely exception of
assumption 1 in several sources) are valid. One probable exception to
assumptions 2 and 7 is the pointlike source LS\,5039, discussed in
detail in~\citet{Aharonian_ls5039}.

%
%
\section{Neutrino event rates} \label{sec:evtRates}
\subsection{Signal event rates} \label{sec:signalRates}
Given a neutrino spectrum $\d N_\nu/\d E_\nu$ at the Earth from a
source the event rate in a neutrino telescope can be calculated as
\begin{eqnarray}
\frac{\d N_{\nu}}{\d t} = 
\int \d E_\nu \ A^\mathrm{eff}_\nu \,\frac{\d N_\nu}{\d E_\nu} \ .
\label{eventrate1}
\end{eqnarray}
Here, $A^\mathrm{eff}_\nu$ is the neutrino effective area of the
detector comprising the detection efficiency of neutrinos with an
Earth-based telescope. Deep-sea neutrino telescopes detect neutrinos
via the measurement of the Cherenkov light emitted by muons produced
in the interaction of high energy neutrinos. Muons produced in
hadronic interactions of charged CR in the Earth's atmosphere present a
background with a flux many orders of magnitude higher than the
expected cosmic neutrino flux. To suppress this background neutrino
telescopes are optimised to observe upward-going neutrinos, using the
Earth as a filter. The neutrino attenuation in the Earth, as well as the
neutrino conversion probability and the muon detection efficiency, are
comprised in the effective area $A^\mathrm{eff}_\nu$.

\begin{figure}[tb]
\begin{center}
\plotone{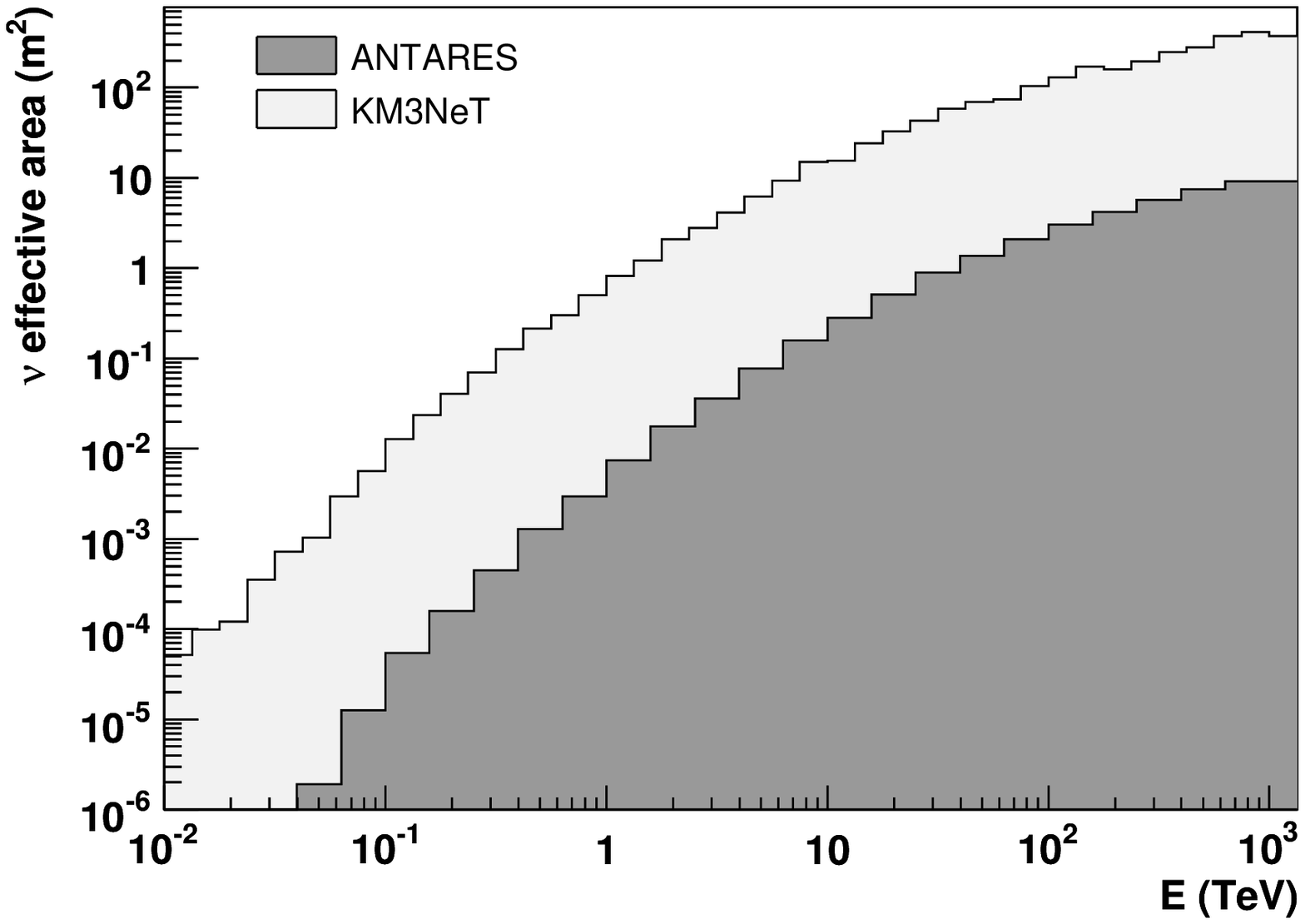}
\caption{ Effective areas $A^\mathrm{eff}_\nu(E_\nu)$ for muon 
neutrinos of the ANTARES detector and the KM3NeT detector with an
instrumented volume of $1 \km^3$ in the energy range from $10 \gev$ to
$1\pev$. }
\label{effectiveArea}
\end{center}
\end{figure}

Due to the increase of the neutrino cross section and the muon range
and its light yield per unit path length with energy the effective
area is energy-dependent. Figure~\ref{effectiveArea} shows the
effective area of the ANTARES detector \citep{effAreaAnt} and the
estimate for a future KM3NeT detector with an instrumented volume of
$1 \km^3$ \citep{effAreaKM3} for muon neutrinos in the energy range
from $10\gev$ to $1\pev$. (Since currently dedicated reconstruction
software for the KM3NeT detector is not yet available, its effective
area is calculated by requiring signals from the muon in at least 10
photo-sensors.) Absorption effects due to the increase of the neutrino
cross section with energy and the resulting increasing opacity of the
Earth start to affect the effective area at about $100\tev$ for small
nadir angles between the detector normal and the source direction and
are thus of no effect in the analysis described here (for the
determination of the event numbers in Tab.\,\ref{tab:rates} the energy
range was limited to $< 100\tev$). Therefore, the effective area is
assumed to be the same for all observation angles between the source
and the detector. Both detectors operate at a neutrino reconstruction
threshold of $\sim$100$\gev$ well matched to the $\gamma$-ray
threshold of H.E.S.S.

The detection of the muon-induced Cherenkov light with a three
dimensional array of photo sensors allows one to reconstruct the
flight direction and energy of the muon, which is used as an estimator
for the flight direction and energy of the primary neutrino. In our
study we neglect the effect of the uncertainty in the energy
determination and use the true neutrino energy instead. The angular
resolution $\sigma_{\mathrm{PSF}}$ of the neutrino direction for a
deep-sea neutrino telescope decreases with increasing energy and is in
the case of ANTARES better than $0.4^\circ$ for neutrino energies
above $1\tev$. For KM3NeT a slightly better resolution can be
expected, where in the energy range considered here the resolution is
dominated by the difference in the flight direction of the neutrino
and its muon. For the sake of simplicity we assume in our study
angular resolutions of $0.4^\circ$ and $0.3^\circ$ for all energies
for the ANTARES and KM3NeT detector, respectively. The optimal search
window for an extended neutrino source, assuming an isotropic
background of atmospheric neutrinos, is then given approximately as
$\Theta_{\mathrm{opt}} = 1.6
\times \sqrt{\sigma_{\mathrm{PSF}}^{2} + \sigma_{\mathrm{src}}^{2}}$,
where $\sigma_{\mathrm{src}}$ is the rms width of the source. This is
also valid for the extended sources RX\,J1713.7$-$3946 and
RX\,J0852.0$-$4622 which exhibit a ring-like region of enhanced
emission. In these two cases convolutions of the point-spread function
of the detector with the emission profiles yield distributions that
can be approximated by Gaussian distributions with $\sigma =
\sqrt{\sigma_{\mathrm{PSF}}^{2} + \sigma_{\mathrm{src}}^{2}}$. For the
diffuse plane emission ($|b|<1^{\circ}$, $|l|<30^{\circ}$) we use a
search window of size $60 \times 2 \, \mathrm{deg}^2$.

The number of observed events with a neutrino telescope is then
calculated by
\begin{eqnarray}
N_{\nu} = R_{\Theta_{\mathrm{opt}}} \int \d t \, \frac{\d N_{\nu}}{\d t}
\label{eventrate2} \ ,
\end{eqnarray}
where the integration runs over the observation time when the source
is below the horizon, and $R_{\Theta_{\mathrm{opt}}} = 0.72$ is a
reduction factor taking into account the loss of source neutrinos
outside the search window (for the diffuse plane emission we use
$R_{\Theta_{\mathrm{opt}}} = 1$).

The differential fluxes of observed source neutrinos from
RX\,J1713.7$-$3946 and Vela\,X are displayed in the middle plots of
Fig.~\ref{spectra} for an observation time of 5\,yr for KM3NeT. The
bottom two plots show the corresponding integrated numbers of events
for neutrino energies $> E_\nu$. Table~\ref{tab:rates} shows the
calculated neutrino event numbers of all Galactic $\gamma$-ray sources
for 5\,yr of observation time with a $1\km^3$ KM3NeT detector for
energies above $1$ and $5\tev$. For each source, the range of possible
neutrino event numbers and the mean neutrino event number (in
parentheses) are displayed. The neutrino event number ranges result
from the $1\, \sigma$ error bands of the fits, taking into account the
systematic uncertainty. For sources with no published claim of a
curvature in the $\gamma$-ray spectrum, the lower bound of the
neutrino event number results from the uncertainty of the fit of a
curved spectrum, whereas the upper bound results from the uncertainty
of the fit of a pure power law. We note that taking a pure power law
instead of a curved spectrum might result in a considerable difference
of expected neutrino rates. Fitting for example a pure power law for
the SNR RX\,J1713.7$-$3946 results in 16.0 neutrinos in 5\,yr for
KM3NeT compared to 10.7 for a curved spectrum (see also the two lower
left plots in Fig.\,\ref{spectra}). For ANTARES, each of the brightest
sources Vela\,X, RX\,J1713.7$-$3946, and RX\,J0852.0$-$4622 yields at
most 0.3 events in 5\,yr. Taking into account the different methods
used, our calculations are in good agreement with recent calculations
of neutrino event rates from Galactic sources from \citet{kistler} and
\citet{distefano}.

\subsection{Background event rates}
Neutrinos produced in hadronic interactions of charged CRs in
the atmosphere (atmospheric neutrinos) on the opposite side of the
Earth result in signals in the telescope indistinguishable from those
of cosmic neutrinos. The atmospheric neutrino event rate can be
calculated in the same way as the source neutrino event rate. The flux
of atmospheric neutrinos rapidly decreases with increasing energy and
also strongly depends on the zenith angle, i.e.\ the time of the
day. For large zenith angles the path length of the pions and muons in
the thin outer atmosphere is larger compared to small zenith angles
resulting in a higher decay probability and thus a larger neutrino
flux. In this study we use the parameterisation of the atmospheric
neutrino flux of \citet{volkova}, which reasonably well parametrises
the energy and zenith angle dependence in the energy range below $100
\tev$. The measured size of the objects in $\Tev$
$\gamma$-rays then allows us to give reliable estimates of the rates
of detection of atmospheric neutrinos from the directions of
$\gamma$-ray production regions.

\begin{figure*}[thb]
\begin{center}
\plottwo{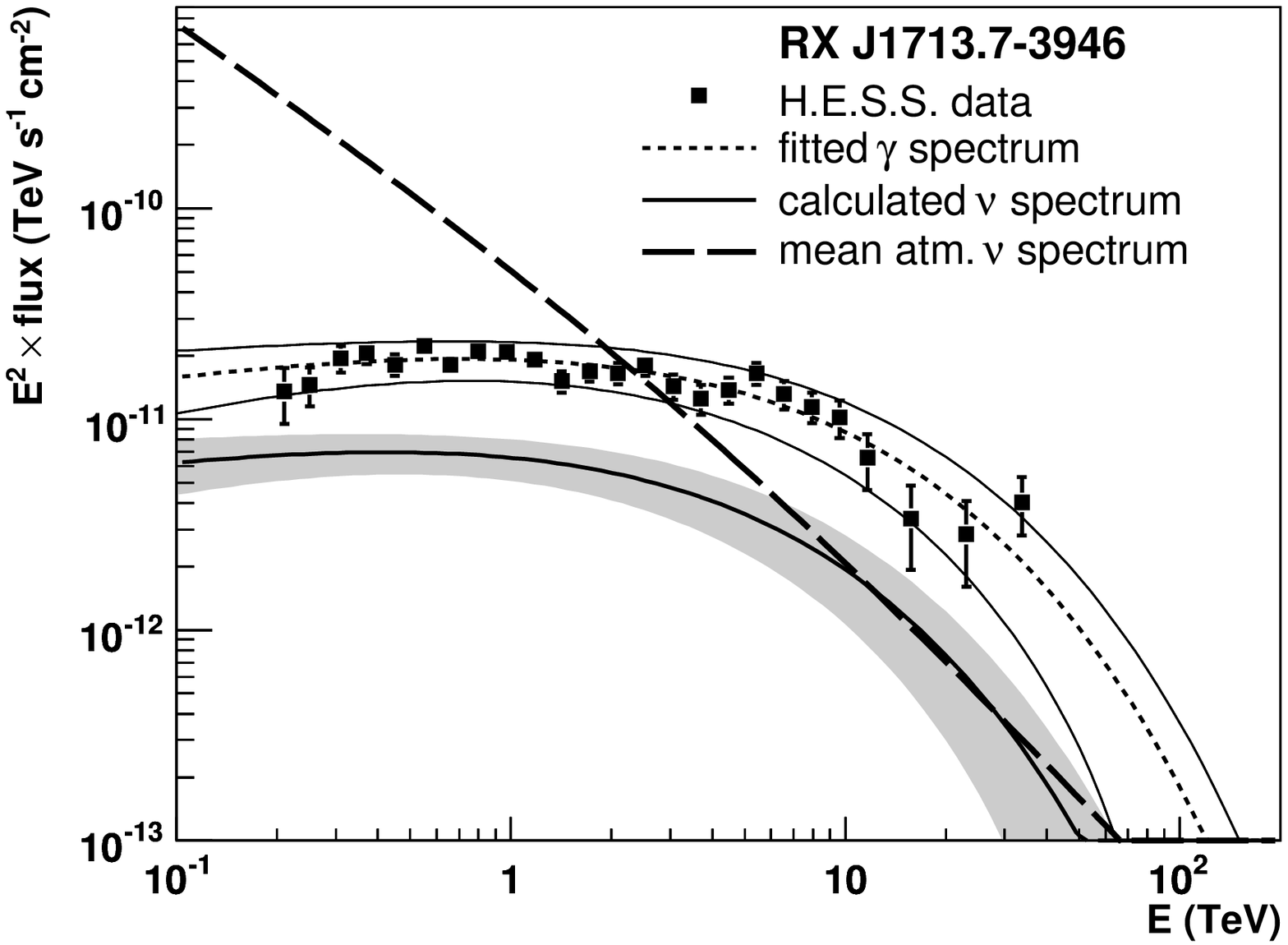}{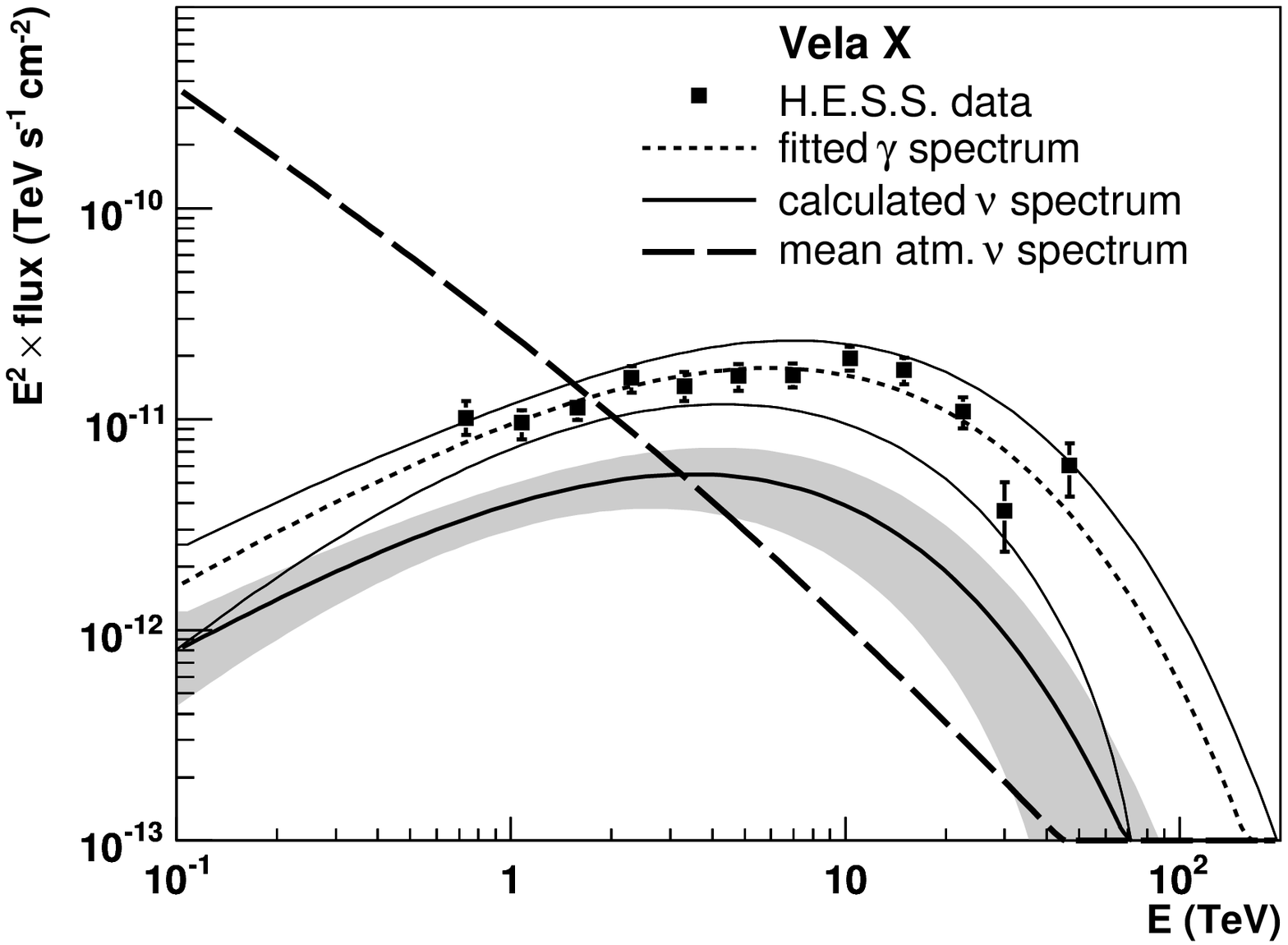}
\plottwo{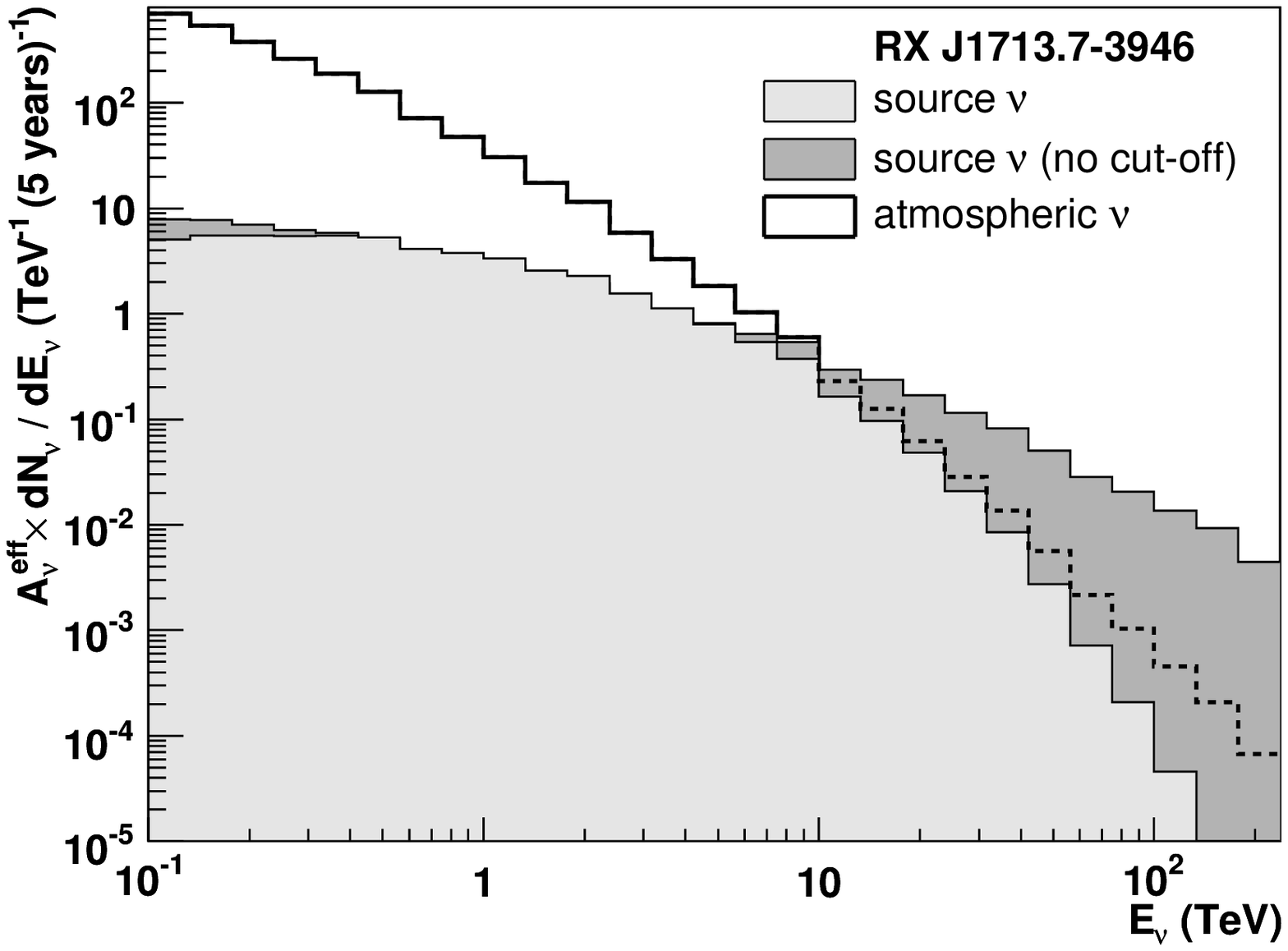}{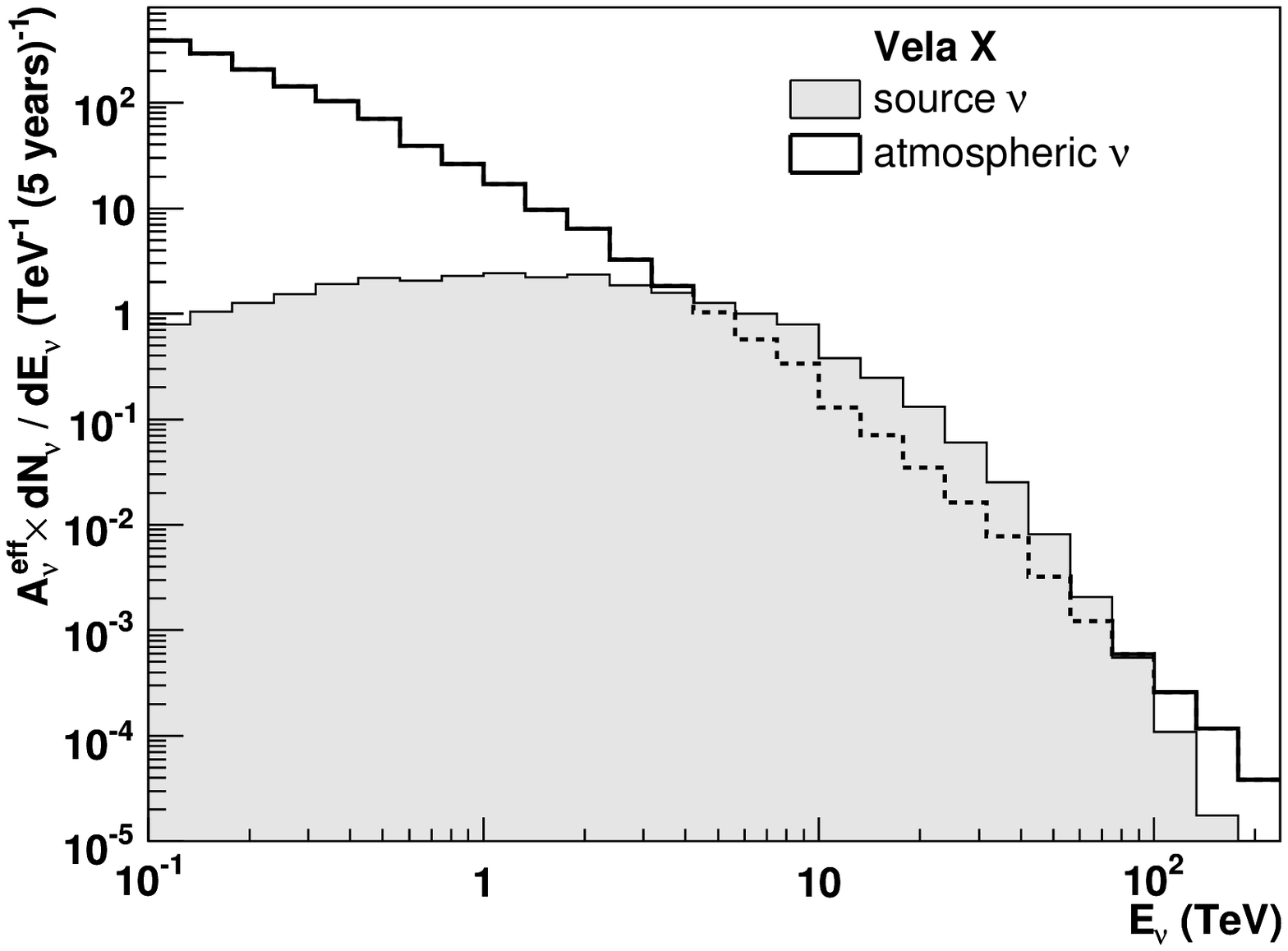}
\plottwo{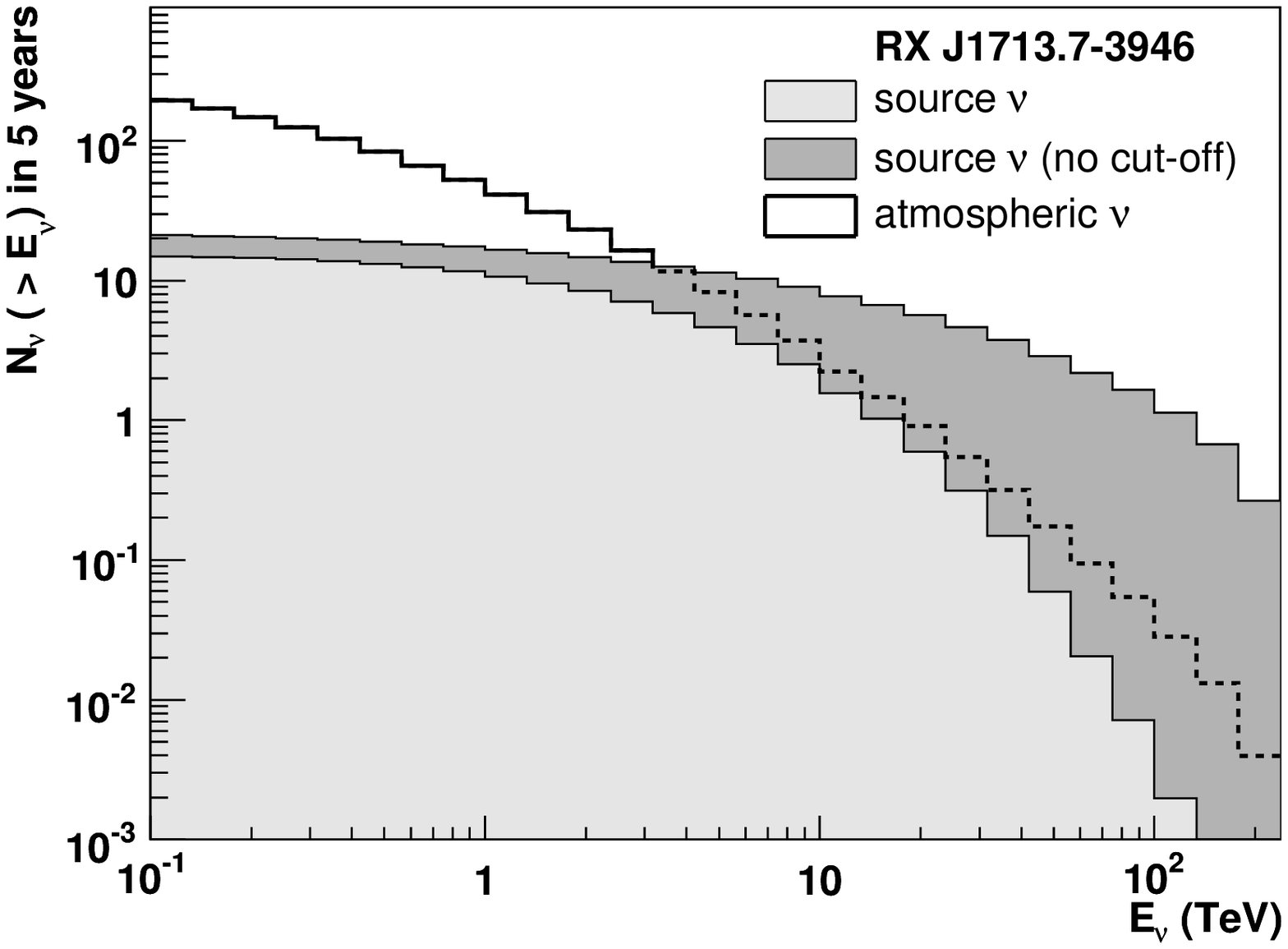}{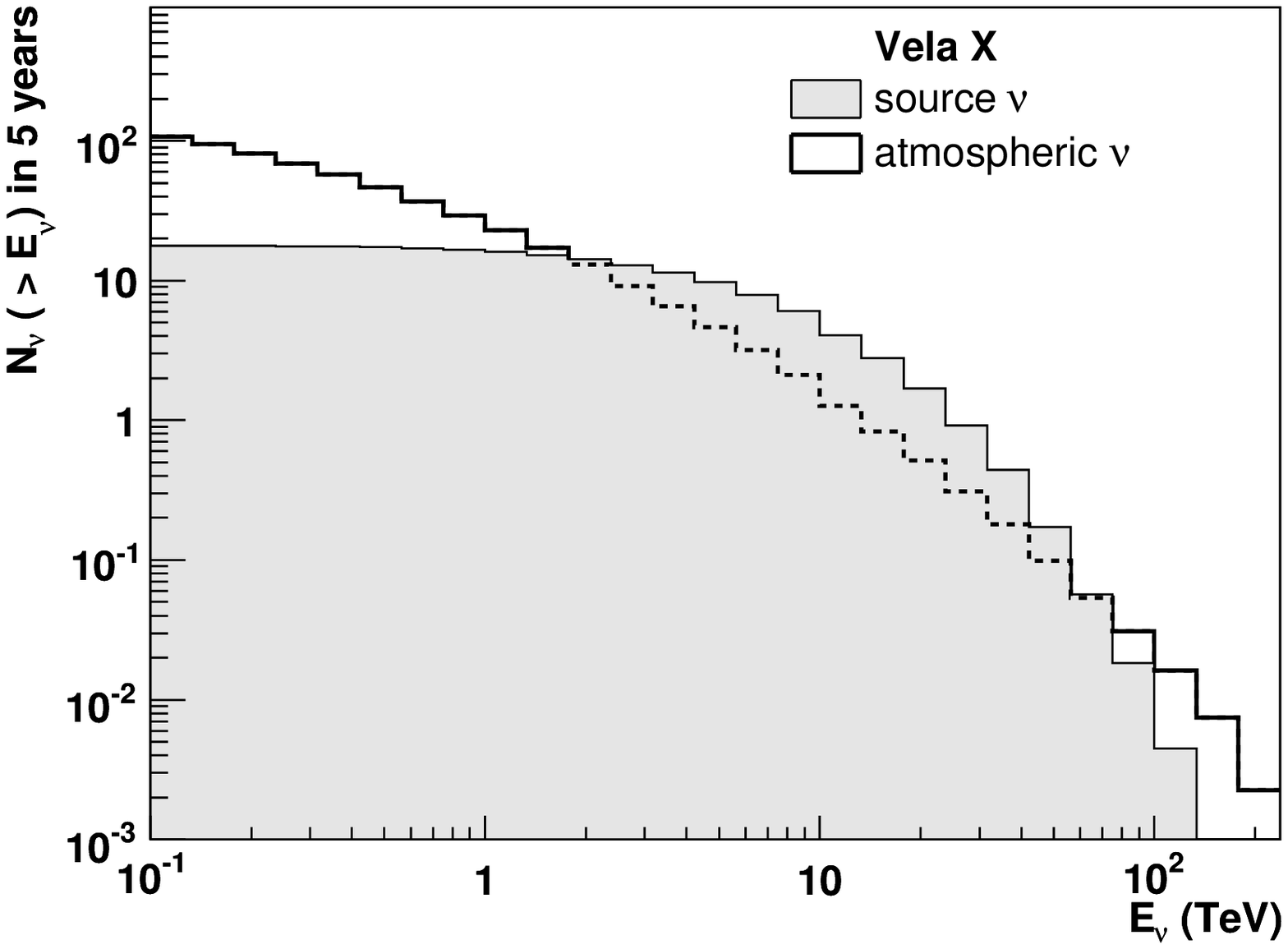}
\caption{\emph{Left}: Plots for RX\,J1713.7$-$3946, \emph{Right}: Plots for
Vela\,X. \emph{Top}: Measured $\gamma$-ray flux and the estimated
neutrino flux with their error bands, together with the atmospheric
neutrino flux. The daily averaged atmospheric neutrino flux is
integrated over the search window. \emph{Middle}: Differential flux of
source neutrinos in 5\,yr, together with the corresponding atmospheric
neutrino flux for KM3NeT. \emph{Bottom}: Number of detected events
with neutrino energies $> E_\nu$ in 5\,yr in the KM3NeT detector. In
addition, for RX\,J1713.7$-$3946 the lower two plots show the case for
a source spectrum assuming no cut-off. In the lower four plots the
limited size of the search window is taken into account, reducing the
number of source neutrinos compared to the upper two plots.}
\label{spectra}
\end{center}
\end{figure*}

The daily averaged atmospheric neutrino fluxes from sky regions of
RX\,J1713.7$-$3946 and Vela\,X integrated over the respective search
windows are shown in Fig.\,\ref{spectra}. Due to the steeper energy
dependence of the atmospheric neutrino flux, the signal-to-background
ratio improves with increasing energy. The calculated event numbers
for atmospheric neutrinos for KM3NeT for 5\,yr are displayed in
Tab.\,\ref{tab:rates}, where we use a mean position $b=0^{\circ}$,
$l=0^{\circ}$ in Galactic coordinates for the diffuse plane
emission. For ANTARES these numbers are at least of the same size as
the maximal numbers of detected source events.

\renewcommand{\arraystretch}{1.2} 
\begin{table*}[thb]
\caption{\label{tab:rates} Shown are H.E.S.S.\ catalog sources with
the range of estimated neutrino event rates and mean rates (in
parentheses) within the search window in KM3NeT for 5 years of
operation together with the expected atmospheric neutrino background
for neutrino energies above $1\tev$ and $5\tev$ (for definition of
range and search window see Sec.\,\ref{sec:evtRates}). The reference
for each source is listed in column \emph{Ref.}\ (all
\emph{H.E.S.S.\ Coll.}). The values under $\varnothing$ represent
the diameter of the $\gamma$-ray source. In the case of non circular
sources, the corresponding dimensions in degrees are given. The column
\emph{Vis.}\ displays the visibility (fraction of time when the source
is below the horizon) of the source to KM3NeT, where for the diffuse
plane emission a mean visibility of 65\% is assumed. The neutrino
spectrum parameters at the Earth, $k_\nu$ and $\epsilon_\nu$, are
given in units of $10^{-12}
\tev^{-1} \cm^{-2} \scnd^{-1}$ and $\Tev$, respectively. A missing
entry for $\epsilon_\nu$ indicates that a pure power law was fitted to
the $\gamma$-ray data points of the respective source. Also, note that
$\Gamma_\nu$ and $\epsilon_\nu$ are strongly correlated.}
\begin{center}
\tiny 
\renewcommand{\arraystretch}{1.2} 
\begin{tabular}{l@{\hspace{2mm}}l@{\hspace{2mm}}c@{}rp{0mm}rrrp{-1mm}  r@{\,--\,}l@{\hspace{1.5mm}}@{}r@{}@{\hspace{0mm}}r@{\hspace*{5mm}}  p{0mm}  r@{\,--\,}l@{\hspace{1.5mm}}@{}r@{}@{}r@{\hspace*{5mm}}} 
\hline\hline 
\multicolumn{4}{c}{} &   &\\[-2mm] 
\multicolumn{4}{c}{} &   & \multicolumn{3}{c}{\bf Spec. parameters} &   &  \multicolumn{4}{c}{\boldmath $E_\nu>1\tev$} &&  \multicolumn{4}{c}{\boldmath $E_\nu>5\tev$} \\ 
\\[-4mm] 
\multicolumn{1}{c}{\bf Source Name} &   \multicolumn{1}{c}{\bf Ref.} &   \multicolumn{1}{c}{\bf \boldmath $\varnothing$ ($^\circ$)} &  \multicolumn{1}{c}{\bf Vis.} &  &  \multicolumn{1}{c}{\bf \boldmath $k_\nu$} &  \multicolumn{1}{c}{\bf \boldmath $\Gamma_\nu$} &  \multicolumn{1}{c}{\bf \boldmath $\epsilon_\nu$} &  &  \multicolumn{3}{c}{\boldmath $N_{\rm{src}}$} &  \multicolumn{1}{c}{\boldmath $N_{\rm{atm}}$} &  &  \multicolumn{3}{c}{\boldmath $N_{\rm{src}}$} &  \multicolumn{1}{c}{\boldmath $N_{\rm{atm}}$} \\ 
 \hline 
 \\[-2mm] 
\multicolumn{18}{c}{\bf Source class A (supernova remnants)} \\ 
 \\[-2mm] 
\hline 
RX\,J1713.7$-$3946&2006d &1.3&0.74&&15.52&1.72&1.35&&7&14&(11) &41&  &2.6 & 6.7&(4.6) &8.2 \\
RX\,J0852.0$-$4622&2007 &2.0&0.83&&16.76&1.78&1.19&&7&15&(11) &104&  &1.9 & 6.5&(4.2) &21 \\
HESS\,J1640$-$465&2006a &0&0.83&&0.93&2.41&&&0.4&3.3&(2.2) &8.7&  &0.0 & 2.1&(1.3) &1.8 \\
HESS\,J1745$-$290\,$\dagger$&2004a &$<0.1$&0.65&&0.85&2.29&&&1.1&2.7&(2.0) &6.4&  &0.5 & 1.8&(1.3) &1.3 \\
HESS\,J1834$-$087&2006a &0.2&0.54&&0.80&2.45&&&0.2&1.7&(1.1) &6.0&  &0.0 & 1.1&(0.7) &1.2 \\
HESS\,J1713$-$381&2006a &0.1&0.73&&0.23&2.28&&&0.0&1.5&(0.6) &7.2&  &0.0 & 1.1&(0.4) &1.4 \\
\\[-3.5mm] 
\multicolumn{9}{l}{Sums for source class A} &\multicolumn{3}{c}{$ \sim 27$} & 173&&\multicolumn{3}{c}{$ \sim 12$} & 35\\ 
\hline 
 \\[-2mm] 
\multicolumn{18}{c}{\bf Source class B (binary systems)} \\ 
 \\[-2mm] 
\hline 
LS\,5039 (INFC)\,$\dagger\dagger$&2006f &0.1&0.57&&2.50&1.61&1.01&&0.3&0.7&(0.5) &2.5&  &0.1 & 0.3&(0.2) &0.5 \\
LS\,5039 (SUPC)\,$\dagger\dagger$&2006f &0.1&0.57&&0.26&2.51&&&0.1&0.3&(0.2) &3.0&  &0.0 & 0.2&(0.1) &0.6 \\
PSR\,B1259$-$63&2005h &$<0.1$&1.00&&0.34&2.72&&&0.1&0.9&(0.6) &9.1&  &0.0 & 0.4&(0.3) &1.7 \\
\hline 
 \\[-2mm] 
\multicolumn{18}{c}{\bf Source class C (no counterparts at other wavelengths)} \\ 
 \\[-2mm] 
\hline 
HESS\,J1303$-$631&2005g &0.3&1.00&&11.99&1.29&0.21&&0.8&2.3&(1.6) &11&  &0.1 & 0.5&(0.3) &2.1 \\
HESS\,J1745$-$303&2006a &0.4&0.66&&1.01&1.79&&&0&18&(9) &9.0&  &0 & 16&(7) &1.8 \\
HESS\,J1614$-$518&2006a &0.5&1.00&&2.41&2.44&&&1&10&(6) &19&  &0.0 & 6.7&(3.7) &4.0 \\
HESS\,J1837$-$069&2006a &0.2&0.53&&1.65&2.27&&&1.2&4.5&(3.3) &5.9&  &0.4 & 3.2&(2.2) &1.2 \\
HESS\,J1634$-$472&2006a &0.2&0.85&&0.64&2.36&&&0.0&3.1&(1.7) &9.8&  &0.0 & 2.2&(1.1) &2.0 \\
HESS\,J1708$-$410&2006a &0.1&0.76&&0.44&2.33&&&0.1&1.6&(1.1) &7.6&  &0.0 & 1.1&(0.7) &1.5 \\
\\[-3.5mm] 
\multicolumn{9}{l}{Sums for source class C} &\multicolumn{3}{c}{$ \sim 23$} & 63&&\multicolumn{3}{c}{$ \sim 15$} & 13\\ 
\hline 
 \\[-2mm] 
\multicolumn{18}{c}{\bf Source class D (pulsar wind nebula)} \\ 
 \\[-2mm] 
\hline 
Vela\,X&2006c &0.8&0.81&&11.75&0.98&0.84&&9&23&(16) &23&  &5 & 15&(10) &4.6 \\
HESS\,J1825$-$137&2006h &0.5&0.57&&10.73&2.08&4.24&&5&10&(8) &9.3&  &2.2 & 5.2&(3.7) &1.8 \\
Crab\,Nebula&2006g &$<0.1$&0.39&&22.38&2.15&1.72&&4.0&7.6&(5.8) &5.2&  &1.1 & 2.7&(1.9) &1.1 \\
HESS\,J1632$-$478&2006a &0.3&0.87&&1.87&2.11&&&0&15&(9) &12&  &0 & 12&(7) &2.4 \\
MSH\,15$-$5{\it 2}&2005d &0.2&1.00&&1.89&2.27&&&3.4&9.6&(7.1) &10&  &1.5 & 6.6&(4.7) &2.0 \\
HESS\,J1616$-$508&2006a &0.3&1.00&&2.11&2.36&&&2.0&9.0&(6.6) &14&  &0.3 & 5.9&(4.1) &3.0 \\
HESS\,J1420$-$607&2006e &0.1&1.00&&1.16&2.25&&&2.0&6.3&(4.6) &9.6&  &0.7 & 4.4&(3.1) &1.9 \\
HESS\,J1418$-$609&2006e &0.1&1.00&&0.94&2.19&&&1.7&6.1&(4.2) &9.6&  &0.8 & 4.5&(3.0) &1.9 \\
HESS\,J1813$-$178&2006a &0.1&0.59&&0.96&2.09&&&0.7&4.6&(3.2) &5.8&  &0.2 & 3.6&(2.4) &1.1 \\
HESS\,J1702$-$420&2006a &0.2&0.77&&0.82&2.32&&&0.5&3.3&(2.1) &8.4&  &0.0 & 2.3&(1.4) &1.7 \\
HESS\,J1804$-$216&2006a &0.4&0.61&&1.49&2.73&&&0.6&2.0&(1.5) &8.4&  &0.1 & 1.0&(0.7) &1.7 \\
G\,0.9+0.1&2005a &$<0.1$&0.65&&0.27&2.31&&&0.1&0.9&(0.6) &6.2&  &0.0 & 0.6&(0.4) &1.2 \\
\\[-3.5mm] 
\multicolumn{9}{l}{Sums for source class D} &\multicolumn{3}{c}{$ \sim 68$} & 122&&\multicolumn{3}{c}{$ \sim 41$} & 24\\ 
\hline 
 \\[-2mm] 
\multicolumn{18}{c}{\bf Diffuse emissions from CR interactions} \\ 
 \\[-2mm] 
\hline 
Gal.\,Centre\,Ridge&2006b &$1.6 \times 0.6$&0.65&&1.29&2.29&&&1.0&4.2&(3.0) &27&  &0.3 & 2.9&(2.0) &5.3 \\
\hline 
 \\[-2mm] 
\multicolumn{18}{c}{\bf Integrated emissions from the Galactic plane} \\ 
 \\[-2mm] 
\hline 
\multicolumn{9}{l}{All known sources} &\multicolumn{3}{c}{$ \sim 122$} & 399&  &\multicolumn{3}{c}{$ \sim 72$} & 80\\ 
\multicolumn{2}{l}{Diffuse\,Plane\,Emission} &$60 \times 2.0$&&&&&&&\multicolumn{3}{c}{24} & 1024 && \multicolumn{3}{c}{11} & 203 \\ 
\hline 
\end{tabular}
\flushleft {$\dagger$ Source association is uncertain; might be associated with the SNR Sgr\,A~East or the Galactic center black hole Sgr\,A$\kern-0.17em^\star$. \\ 
$\dagger\dagger$ Assuming no $\gamma$-ray absorption within the source. INFC and SUPC specify the two phases of inferior and superior conjunction of the binary system as defined in \citet{HESS_ls5039aa}.} 

\end{center}    
\end{table*}

%
%
\section{Discussion}

Sensitive $\Tev$ $\gamma$-ray detectors have surveyed most of our
Galaxy, and it is likely that all of the bright Galactic $\gamma$-ray
sources have now been identified. Most of these bright sources are
located in the southern sky and are hence best visible (at energies
below $\sim 100 \tev$) to neutrino telescopes located in the Northern
Hemisphere.

Under the assumption that the $\gamma$-rays are dominantly produced
via $p$-$p$ interactions we use $\gamma$-ray spectra measured with
H.E.S.S.\ to derive the associated neutrino emission. Our neutrino
flux calculation differs from previous work in two important respects:
the use of new parametrisations of the $\gamma$ and neutrino
production in $p$-$p$ interactions and the use of \emph{curved}
$\gamma$-ray source spectra if such a claim is published.

We have calculated the expected neutrino event rates for two neutrino
telescopes in the Mediterranean Sea: ANTARES (currently under
construction) and KM3NeT, a future km$^{3}$ scale detector. For the
calculations we used neutrino telescope effective areas based on full
simulations of the detector response and took into account the effect
of optimal search windows.

We find that the brightest $\gamma$-ray sources produce neutrino rates
above $1\tev$, comparable to the background from atmospheric
neutrinos. The expected event rates of these sources in the ANTARES
detector make a detection unlikely. For KM3NeT, with an instrumented
volume of $1 \km^3$, an event rate of a few neutrinos per year ($E_\nu
> 1 \tev$) from each of the three brightest $\gamma$-ray sources
(RX\,J1713.7$-$3946, Vela\,X, and RX\,J0852$-$4622) can be expected,
and the detection of individual sources seems possible after several
years of stable detector operation (see Table
\ref{tab:rates}). However, because of the low statistics of source
neutrinos the detection of $\Tev$ signals from a major fraction of
H.E.S.S.\ sources will be difficult for $1 \km^3$ class neutrino
telescopes.

We would like to point out that the event rates presented in this
paper are based on preliminary calculations of the KM3NeT effective
neutrino area, which, e.g., do not include the muon reconstruction
procedure (see Sec.\,\ref{sec:signalRates}). Therefore, future more
detailed calculations will probably lead to a reduction of the
effective area. Also, we want to note that the estimated rates should
be treated as upper limits for most sources, since a significant
contribution from directly accelerated electrons to the $\gamma$-ray
fluxes is possible (or likely). On the other hand, the neutrino rates
from sources in which $\gamma$-rays are heavily absorbed could be
significantly (up to orders of magnitude)
\emph{higher} than those given here. X-ray binaries represent one
source class in which such absorption is likely.

When this paper was completed we learned that \citet{vissani}
published a calculation of neutrino event rates for RX\,J1713.7$-$3946
in which he also uses a primary spectrum with a cut-off at high
energies. The quoted event numbers are in good agreement with those
given in this paper taking into account the differences in the two
analyses.

%
%
\acknowledgments
J.~H. acknowledges the support of the German BMBF through
Verbundforschung Astro-Teilchenphysik (05CH5VH1/0).

%
%

\end{document}